\documentclass[conference,letterpaper]{IEEEtran}
\usepackage[utf8]{inputenc}
\usepackage[T1]{fontenc}
\usepackage{url}
\usepackage{ifthen}
\usepackage{cite}
\usepackage[cmex10]{amsmath} 
\usepackage{graphicx}
\usepackage{amsfonts}
\usepackage{amssymb}
\usepackage{mathrsfs}
\usepackage{algorithm}

\newtheorem{example}{\textbf{Example}}

\begin{document}
\title{Statistical Learning Aided Decoding of BMST of Tail-Biting Convolutional Code}



 \author{%
   \IEEEauthorblockN{Xiao Ma\IEEEauthorrefmark{1},
                     Wenchao Lin\IEEEauthorrefmark{1},
                     Suihua Cai\IEEEauthorrefmark{2},
                     and Baodian Wei\IEEEauthorrefmark{1}}
   \IEEEauthorblockA{\IEEEauthorrefmark{1}School of Data and Computer Science, Sun Yat-sen University, Guangzhou 510006, China\\
                     \IEEEauthorrefmark{2}School of Electronics and Information Technology, Sun Yat-sen University, Guangzhou 510006, China\\
                     Email: maxiao@mail.sysu.edu.cn, linwch7@mail2.sysu.edu.cn, caish5@mail2.sysu.edu.cn, weibd@mail.sysu.edu.cn}
 }


\maketitle

\begin{abstract}
  This paper is concerned with block Markov superposition transmission~(BMST) of tail-biting convolutional code~(TBCC). We propose a new decoding algorithm for BMST-TBCC, which integrates a serial list Viterbi algorithm~(SLVA) with a soft check instead of conventional cyclic redundancy check~(CRC). The basic idea is that, compared with an erroneous candidate codeword, the correct candidate codeword for the first sub-frame has less influence on the output of Viterbi algorithm for the second sub-frame. The threshold is then determined by statistical learning based on the introduced empirical divergence function. The numerical results illustrate that, under the constraint of equivalent decoding delay, the BMST-TBCC has comparable performance with the polar codes. As a result, BMST-TBCCs may find applications in the scenarios of the streaming ultra-reliable and low latency communication~(URLLC) data services.
\end{abstract}

\begin{IEEEkeywords}
Block Markov superposition transmission~(BMST), list decoding, statistical learning, ultra-reliable and low latency communication~(URLLC).
\end{IEEEkeywords}

\section{Introduction}
It has been pointed out by Shannon~\cite{Shannon1948Theory} that the error free transmission is possible with infinite coding length as long as the transmission rate is below the channel capacity. To approach the channel capacity, a number of powerful iteratively decodable channel codes with long block length have been proposed. For example, low-density parity check~(LDPC) codes~\cite{Gallager1962ldpc} and turbo codes~\cite{Berrou1993turbo} perform within a few hundredths of a decibel from the Shannon limits under iterative belief propagation~(BP) decoding algorithm. However, long codes are not suitable for emerging applications that are sensitive to the delay, such as automated driving, smart grids, industrial automation and medical applications. Designing a good code with strict latency constraint is challenging since most constructions developed for long block length do not deliver good codes in the short block length regime. For this reason, more attention has been paid recently on the design of short and medium block length codes~(e.g., a thousand or less information bits)~\cite{Wonterghem2016EBCH}.

One solution is to construct LDPC codes by progressive edge growth~(PEG) algorithm~\cite{Hu2005PEG}, which can deliver better codes than randomly constructed LDPC codes in short block length regime. Polar codes~\cite{Arikan2009Polar}, another promising solution for short packet transmission, have been adopted by 5G control channel. Powerful classical short codes with \text{near} maximum likelihood~(ML) decoding algorithm was also investigated for low latency communication. In~\cite{Wonterghem2016EBCH}, the extended Bose-Chaudhuri-Hocquenghem codes were shown to perform near the normal approximation benchmark under ordered statistics decoding~\cite{Fossorier1995OSD}. As shown in~\cite{Gaudio2017TBCC}, in the short block length regime, the tail-biting convolutional codes~(TBCCs) with wrap-around Viterbi algorithm~\cite{Shao2003WAVA} outperform significantly state-of-the-art iterative coding schemes.

All the aforementioned codes are block codes with short coding length, whereas convolutional codes with limited decoding window can be alternative choices for the streaming services with strict latency constraint, such as real-time online games and video conference. The comparison in~\cite{Rachinger2015ComparisonCCLDPC}~\cite{Maiya2012ComparisonCCLDPC} between convolutional codes and PEG-LDPC codes showed that convolutional codes outperform LDPC codes for very short delay when bit error rate is used as a performance metric.

A coding scheme called block Markov superposition transmission~(BMST) was proposed in~\cite{Ma2015Block} to construct iteratively decodable convolutional codes with long constraint length from simple basic codes. The construction of BMST codes is flexible, in the sense that it applies to all code rates of interest in the interval $(0,1)$~\cite{Liang2016RUN} and is capable of supporting a wide range of delays but with a small amount of extra implementation complexity~\cite{Zhao2016DelayTunable}. The extrinsic information transfer~(EXIT) chart analysis in~\cite{Huang2016EXIT} showed that BMST codes have near-capacity performance in the waterfall region and an error floor that can be controlled by the encoding memory. However, even with the sliding window decoding algorithm, the BMST codes still suffer from a large decoding delay, which renders BMST codes unsuitable for low latency communication. This is because the BP decoding algorithm performs far worse than the optimal decoding algorithm when the layers~(sub-blocks) become short.

To solve this issue, the semi-random block oriented convolutional code~(SRBO-CC) was proposed in~\cite{Ma2018SRBOCC} with a Cartesian product of short code as the basic code. The SRBO-CC can be decoded by the sequential decoding, whose memory load is heavy due to the requirement of a large amount of stack memory. In~\cite{Ma2019List}, taking the truncated convolutional code as the basic code, we proposed a list decoding algorithm for SRBO-CC. However, the frame error rate of short convolutional codes without termination is relatively high. In this paper, we take a powerful short TBCC as the basic code to build a BMST-TBCC system, where the random interleaver is replaced by a {\em totally random} linear transformation. The BMST-TBCC can be decoded by a successive cancellation decoding algorithm, whose performance depends critically on the performance of the first sub-frame. To recover the first sub-frame reliably, list decoding is conducted and the transmitted codeword is identified from the list with the help of the empirical divergence function. Simulation results show that the BMST-TBCC with successive cancellation decoding algorithm is competitive with the polar codes and that the performance-complexity tradeoffs can be achieved by adjusting the statistical threshold.


\section{BMST of Tail-biting Convolutional Code}\label{SEC_2}
\subsection{Encoding}
\begin{figure}
  \centering
  \includegraphics[width=6.5cm]{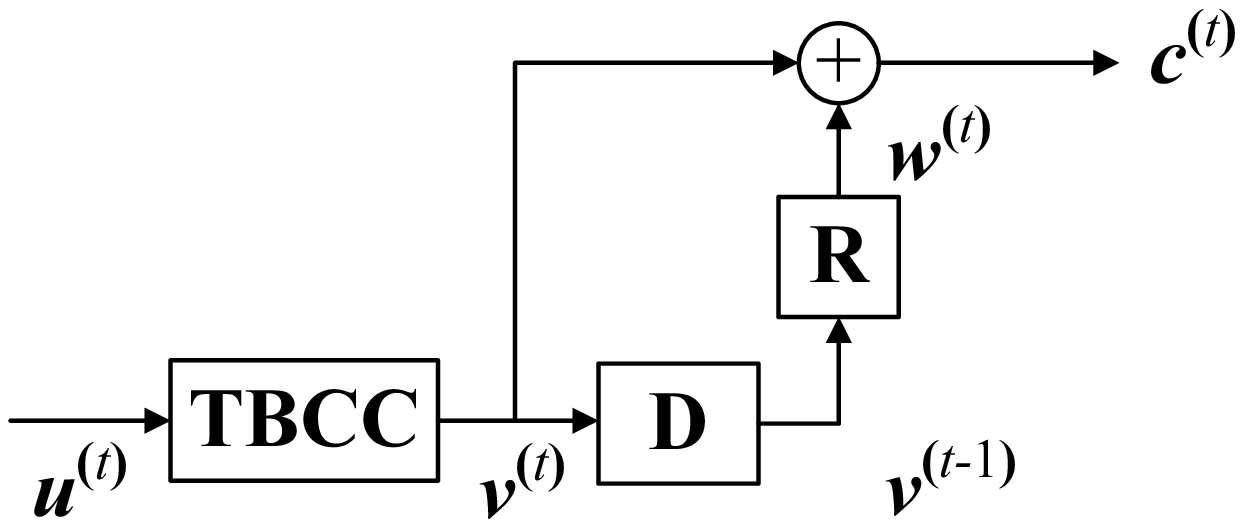}\\
  \caption{Encoding structure of BMST-TBCC system.}\label{FIG_ENC}
\end{figure}

Let $\boldsymbol{u} = (\boldsymbol{u}^{(0)}, \boldsymbol{u}^{(1)},\cdots,\boldsymbol{u}^{(L-1)})$ be the data to be transmitted, where $\boldsymbol{u}^{(t)}=(u^{(t)}_0,u^{(t)}_1,\cdots,u^{(t)}_{k-1})\in \mathbb{F}_2^k$ for $0\leqslant t \leqslant L-1$. The encoding algorithm of BMST-TBCC with basic code $\mathscr{C}$ is described in Algorithm~\ref{EncodingAlgorithm}~(see Fig.~\ref{FIG_ENC} for reference). We see that the main difference from the conventional BMST is the replacement of the random interleaver in the original BMST with a totally random linear transformer $\mathbf{R}$. Also note that we focus on the case with encoding memory one to minimize the rate loss due to the termination.

\begin{algorithm}\caption{Encoding of BMST-TBCC}\label{EncodingAlgorithm}
\begin{itemize}
  \item{\textbf{Initialization:}} \label{step:encoding_initialize} Let $\boldsymbol{v}^{(-1)} = \boldsymbol{0} \in \mathbb{F}_2^{n}$.
  \item{\textbf{Iteration:}} \label{step:encoding_loop} For $0\leqslant t \leqslant L-1$,
      \begin{enumerate}
        \item Encode $\boldsymbol{u}^{(t)}$ into $\boldsymbol{v}^{(t)}\in \mathbb{F}_2^n$ by the encoding algorithm of the basic code $\mathscr{C}$.
        \item Compute $\boldsymbol{w}^{(t)} = \boldsymbol{v}^{(t-1)}\mathbf{R}\in \mathbb{F}_2^n$, where $\mathbf{R}$ is a random matrix of order $n$ whose elements are generated independently according to the Bernoulli distribution with success probability $1/2$.
        \item Compute $\boldsymbol{c}^{(t)} = \boldsymbol{v}^{(t)}+\boldsymbol{w}^{(t)}\in \mathbb{F}_2^{n}$, which will be taken as the $t$-th sub-frame for transmission.
      \end{enumerate}
  \item{\textbf{Termination:}} \label{step:encoding_termination}
        The $L$-th sub-frame is set to $\boldsymbol{c}^{(L)} = \boldsymbol{v}^{(L-1)}\mathbf{R}$, which is equivalent to setting $\boldsymbol{u}^{(L)} = \boldsymbol{0}$.
\end{itemize}
\end{algorithm}

\subsection{Performance Metric}
Suppose that $\boldsymbol{c}^{(t)}$ is modulated with binary phase-shift keying~(BPSK) signals and transmitted over additive white Gaussian noise~(AWGN) channels, resulting in a noisy version $\boldsymbol{y}^{(t)}\in \mathbb{R}^n$ at the receiver. We focus on a sliding window decoding algorithm, which attempts to recover $\boldsymbol{u}^{(t)}$ from $\boldsymbol{y}^{(t)}$ and $\boldsymbol{y}^{(t+1)}$. In other words, the decoding window is two and hence the decoding delay is $2n$. Given a decoding algorithm, define subFER as the probability that the decoding result $\hat{\boldsymbol{u}}^{(0)}$ is not equal to the transmitted vector $\boldsymbol{u}^{(0)}$ and FER as the probability that the decoding result $\hat{\boldsymbol{u}}$ is not equal to $\boldsymbol{u}$. Clearly, we have
\begin{equation}\label{bound}
  {\rm subFER} \leqslant {\rm FER} \leqslant L\cdot {\rm subFER}.
\end{equation}
In practice, we define
\begin{equation}\label{fER}
  {\rm fER} = \frac{\textrm{number of erroneous decoded sub-frames}} {\textrm{total number of transmitted sub-frames}}.
\end{equation}
The event that the decoding result $\hat{\boldsymbol{u}}^{(0)}$ is not equal to the transmitted vector $\boldsymbol{u}^{(0)}$  is referred to as the {\em first error event}. In general, we say that the first error event occurs at time $t$ if $\hat{\boldsymbol{u}}^{(i)} = \boldsymbol{u}^{(i)}$ for all $i < t$ but $\hat{\boldsymbol{u}}^{(t)} \neq \boldsymbol{u}^{(t)}$. Taking into account that the first error event may cause catastrophic error-propagation, we can prove~(omitted here) that
\begin{equation}
{\rm fER} \lessapprox \frac{L}{2} \cdot {\rm subFER}.
\end{equation}

\subsection{Weight Enumerating Function}
We see that the performance is closely related to the subFER, which, in turn, is closely related to the weight distribution of the truncated code $\mathscr{C}^{(0,1)} = \{(\boldsymbol{c}^{(0)}, \boldsymbol{c}^{(1)}) | \boldsymbol{c}=(\boldsymbol{c}^{(0)}, \cdots, \boldsymbol{c}^{(L)}) \; \textrm{is a coded sequence with}\; \boldsymbol{c}^{(0)} \neq \boldsymbol{0} \}$. Let $A(X)$ be the weight enumerating function of the basic code $\mathscr{C}\backslash{\boldsymbol{0}}$~(all non-zero codewords). Then the {\em ensemble} weight enumerating function of the truncated code $\mathscr{C}^{(0,1)}$ with $\mathbf{R}$ being totally random  is given by
\begin{equation}
B(X) = 2^{-n+k}(1+X)^n A(X),
\end{equation}
which can be used to upper-bound subFER if maximum likelihood decoding of $\boldsymbol{c}^{(0)}$ could be implemented based on $(\boldsymbol{y}^{(0)}, \boldsymbol{y}^{(1)})$.

\section{Off-Line Statistical Learning}\label{SEC_3}
\subsection{List Decoding}

We assume that the basic code $\mathscr{C}$ can be efficiently decoded by outputting a list of candidate codewords. To avoid messy notation, we assume that a codeword $\boldsymbol{v}\in \mathscr{C}$ is transmitted. Upon receiving its noisy version $\boldsymbol{y}= (y_0, y_1, \cdots, y_{n-1})$, the decoder {\em serially} outputs a list of candidate codewords $\hat{\boldsymbol{v}}_{\ell}$, $\ell = 1, 2, \cdots, \ell_{\max}$, where $\ell_{\max}$ is a parameter to trade off the performance against the complexity. We will not focus on the detailed implementation in this paper but simply conduct the serial list Viterbi algorithm~(SLVA)~\cite{Seshadri1994LVA} with the tail-biting constraint. For ease of notation, we use SLVA($\boldsymbol{y}$, $\ell$) to represent the $\ell$-th output of the SLVA. In particular, SLVA($\boldsymbol{y}$, 1), simply denoted by VA($\boldsymbol{y}$), is the output of the Viterbi algorithm~(VA).

For any binary vector $\boldsymbol{x}$, its likelihood is given by $f(\boldsymbol{y}|\boldsymbol{x}) = \prod_{i=0}^{n-1} f(y_i|x_i)$, where $f(y_i|x_i)$ is the considered conditional probability density function specified by the modulation and the channel. By the nature of the SLVA, we have $f(\boldsymbol{y}|\hat{\boldsymbol{v}}_1) \geqslant  f(\boldsymbol{y}|\hat{\boldsymbol{v}}_2) \geqslant \cdots \geqslant f(\boldsymbol{y}|\hat{\boldsymbol{v}}_{\ell_{\max}})$, where $\hat{\boldsymbol{v}}_\ell$=SLVA($\boldsymbol{y}$, $\ell$). The list decoding is {\em successful} if the transmitted one is included in the list. Obviously, the probability of the list decoding being successful can be close to one by enlarging the list size $\ell_{\max}$. \textbf{Example}~\ref{EX_LIST} shows the performance of TBCC under list decoding.

\begin{figure}
  \centering
  \includegraphics[width=8cm]{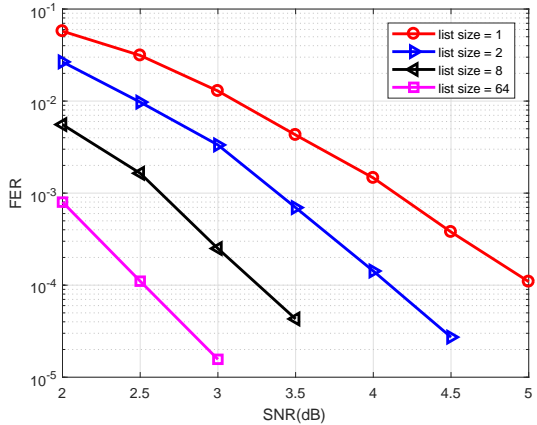}\\
  \caption{Performance of TBCC under list decoding.}\label{FIG_LIST}
\end{figure}

\begin{example}\label{EX_LIST}
The $16$-state $(2,1,4)$ TBCC defined by the polynomial generator matrix $G(D) = [10111, 11001]$ with information length $k=32$~($n=64$) is considered. The list decoding performance is shown in Fig.~\ref{FIG_LIST}, where we observe that the performance can be improved by increasing the list size.
\end{example}

For a large list size~(e.g., $\ell_{\rm max}=64$), the transmitted codeword is included in the list with high probability. Then the key issue is how to identify the transmitted one from the list. One solution is to invoke the cyclic redundancy check~(CRC), as embedded in polar codes~\cite{Niu2012CRCPolar}. However, the overhead~(rate loss) due to the CRC is intolerable especially for short TBCCs. This motivates us to employ the intrinsic memory of the BMST system. The basic observation is that, compared with an erroneous candidate codeword, the correct candidate codeword for the first sub-frame has less influence on the output of Viterbi algorithm for the second sub-frame. To proceed, we need the following concept.

\subsection{Empirical Divergence Function}
For the received vector $\boldsymbol{y}\in \mathbb{R}^n$, define an {\em empirical divergence function}~(EDF) as
\begin{equation}
D(\boldsymbol{x}, \boldsymbol{y}) = \frac{1}{n} \log_2 \frac{f(\boldsymbol{y}|\boldsymbol{x})}{f(\boldsymbol{y})}
\end{equation}
for $\boldsymbol{x} \in \mathbb{F}_2^n$, where
\begin{equation}
f(\boldsymbol{y}) = \prod_{i=0}^{n-1} \left(\frac{1}{2}f(y_i|0)+\frac{1}{2}f(y_i|1)\right).
\end{equation}
Note that, in the above definition,  $f(\boldsymbol{y})$ is not equal to $2^{-k}\sum_{\boldsymbol{v}\in \mathscr{C}}f(\boldsymbol{y}|\boldsymbol{v})$ but to $2^{-n}\sum_{\boldsymbol{x}\in \mathbb{F}_2^n}f(\boldsymbol{y}|\boldsymbol{x})$. Also note that the vector $\boldsymbol{y}$ is not necessarily  the noisy version of $\boldsymbol{x}$. We are interested in the following cases.

\begin{enumerate}
  \item If $\boldsymbol{v}$ is the transmitted one, we have $D(\boldsymbol{v}, \boldsymbol{y})\approx I(X; Y) > 0$, where $\approx$ is used to indicate that the EDF is around in probability its expectation for large $n$. Here $I(X; Y)$ is the mutual information between the channel output $Y$ and the uniform binary input $X$.
  \item If $\boldsymbol{x}$ is randomly generated~(hence typically not equal to the transmitted one), we have
      \begin{equation}
      D(\boldsymbol{x}, \!\boldsymbol{y})\!\approx\! \mathbb{E}_{Y\!|\!V}\!\left[\frac{1}{2}\log_2\!\frac{f(Y|0)}{f(Y)}\!+\!\frac{1}{2}\log_2\!\frac{f(Y|1)}{f(Y)} \right],
      \end{equation}
      which is negative from the concavity of the function $\log_2(\cdot)$.
  \item What are the typical values of $D(\hat{\boldsymbol{v}}, \boldsymbol{y})$, where $\hat{\boldsymbol{v}}={\rm VA}(\boldsymbol{y})$? Given $\boldsymbol{y}$, since $D(\hat{\boldsymbol{v}}, \boldsymbol{y}) = \max_{\boldsymbol{v}\in \mathscr{C}}D(\boldsymbol{v},\boldsymbol{y})$, we expect that $D(\hat{\boldsymbol{v}}, \boldsymbol{y})\gtrapprox I(X; Y) > 0$.
  \item What about $D(\tilde{\boldsymbol{v}}, \tilde{\boldsymbol{y}})$? Here $\tilde{\boldsymbol{v}} = {\rm VA}(\tilde{\boldsymbol{y}})$ where $\tilde{\boldsymbol{y}} = \boldsymbol{x} \odot \boldsymbol{y}$ with $\boldsymbol{x}$ being a totally random bipolar vector, where $\odot$ stands for component-wise product. That is, we first randomly flip the received vector, and then execute the VA to find the first candidate codeword $\tilde{\boldsymbol{v}}$. We expect that $D(\tilde{\boldsymbol{v}}, \tilde{\boldsymbol{y}})$ is located between $D(\boldsymbol{v}, \boldsymbol{y})$ of the first case and $D(\boldsymbol{x}, \boldsymbol{y})$ of the second case.

\end{enumerate}

\begin{figure}
  \centering
  \includegraphics[width=8cm]{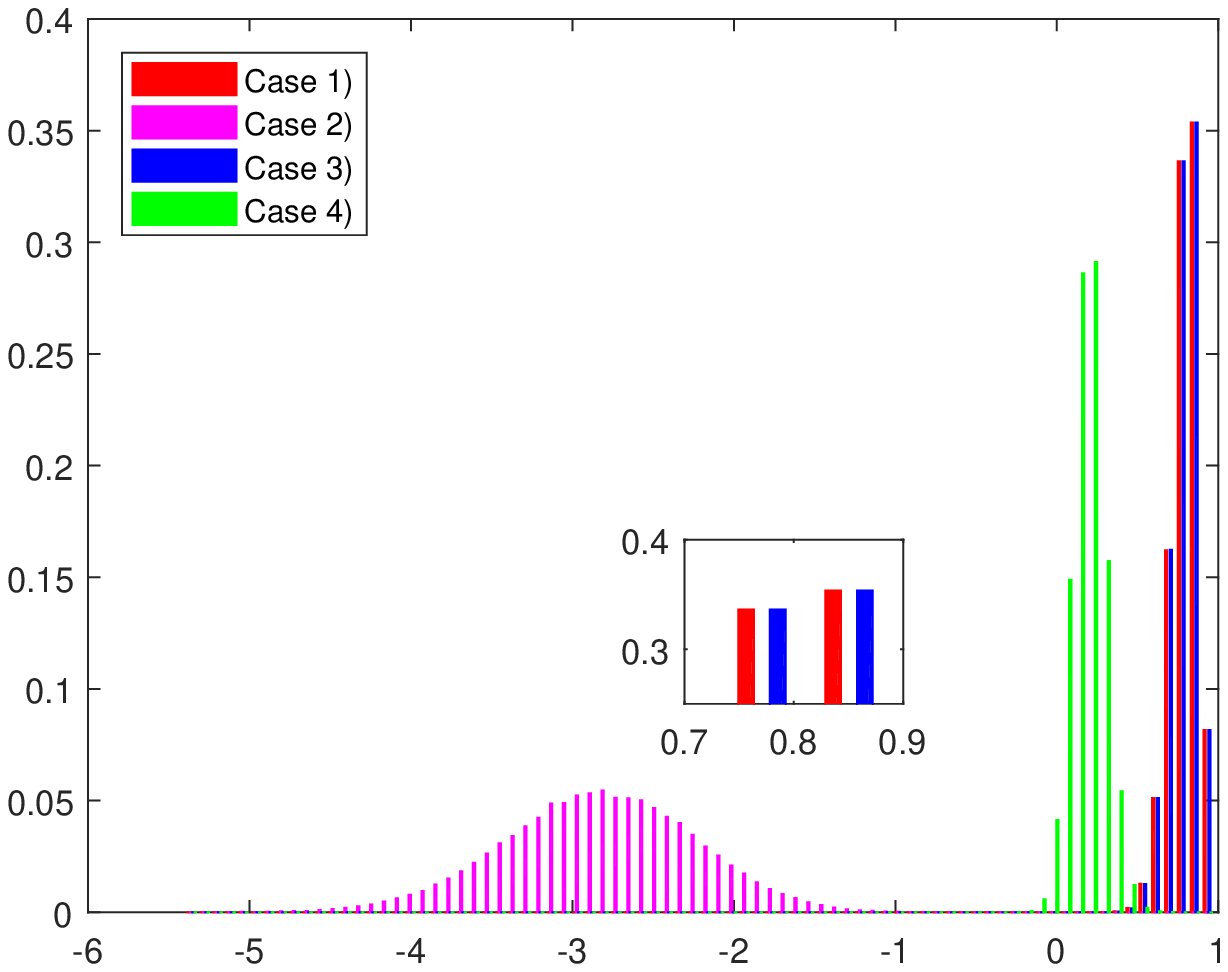}\\
  \caption{Statistical behavior of $D(\boldsymbol{x},\boldsymbol{y})$.}\label{FIG_STATISTIC}
\end{figure}

\begin{example}\label{EX_STATISTIC}
Consider the TBCC in \textbf{Example}~\ref{EX_LIST} again and set ${\rm SNR} = 4~{\rm dB}$, at which the mutual information is $I(X; Y)\approx0.79$. The histogram is shown in Fig.~\ref{FIG_STATISTIC}, from which we observed that $D(\boldsymbol{v}, \boldsymbol{y})$ is likely to be large with $\boldsymbol{v}$ being the transmitted one~(or the output of the VA corresponding to $\boldsymbol{y}$). Note that the statistical behavior of $D(\tilde{\boldsymbol{v}}, \tilde{\boldsymbol{y}})$ is different from that of $D(\boldsymbol{x}, \boldsymbol{y})$, since $\tilde{\boldsymbol{v}}$ is dependent on $\tilde{\boldsymbol{y}}$. The typical values of $D(\tilde{\boldsymbol{v}}, \tilde{\boldsymbol{y}})$ are greater than those of $D(\boldsymbol{x}, \boldsymbol{y})$ but less than those of $D(\boldsymbol{v}, \boldsymbol{y})$.
\end{example}

The statistical behavior of the EDF can be helpful in the decoding process of BMST-TBCC. In the case when the decoding result of the first sub-frame $\hat{\boldsymbol{v}}^{(0)}=\boldsymbol{v}^{(0)}$, $\boldsymbol{y}^{(1)}\odot\phi(\hat{\boldsymbol{v}}^{(0)}\mathbf{R})$ is the Gaussian noisy version of $\boldsymbol{v}^{(1)}$, where $\phi(\hat{\boldsymbol{v}}^{(0)}\mathbf{R})$ is the BPSK signal corresponding to the binary vector $\hat{\boldsymbol{v}}^{(0)}\mathbf{R}$. In contrast, in the case when $\hat{\boldsymbol{v}}^{(0)}\neq\boldsymbol{v}^{(0)}$, $\boldsymbol{y}^{(1)}\odot\phi(\hat{\boldsymbol{v}}^{(0)}\mathbf{R})$ is the randomly flipped Gaussian noisy version of $\boldsymbol{v}^{(1)}$. Since these two cases have different statistical impact on the EDF, we are able to distinguish with high probability whether $\boldsymbol{y}^{(1)}\odot\phi(\hat{\boldsymbol{v}}^{(0)}\mathbf{R})$ is randomly flipped~(equivalently, $\hat{\boldsymbol{v}}^{(0)}$ is erroneous) or not.

\section{Successive Cancellation Decoding Algorithm}\label{SEC_4}

The BMST-TBCC can be decoded by a sliding window algorithm with successive cancellation, and the critical step is how to recover reliably $\boldsymbol{v}^{(0)}$, which is not interfered by any other sub-frames. In this section, we propose a method to estimate $\boldsymbol{v}^{(0)}$ from $\boldsymbol{y}^{(0)}$ and $\boldsymbol{y}^{(1)}$.

Given $\boldsymbol{y}^{(0)}$, the SLVA is implemented to deliver serially a list of candidates $\hat{\boldsymbol{v}}^{(0)}_\ell, 1\leqslant \ell \leqslant \ell_{\rm max}$. For each candidate codeword, we define a soft metric
\begin{equation}\label{SOFT_METRIC}
M(\hat{\boldsymbol{v}}^{(0)}_\ell) = D(\hat{\boldsymbol{v}}^{(0)}_\ell,\boldsymbol{y}^{(0)}) + D(\tilde{\boldsymbol{v}}_\ell,\boldsymbol{y}^{(1)}\odot\phi(\hat{\boldsymbol{v}}^{(0)}_\ell\mathbf{R})),
\end{equation}
where $\tilde{\boldsymbol{v}}_\ell$ is the output of the VA with $\boldsymbol{y}^{(1)}\odot\phi(\hat{\boldsymbol{v}}^{(0)}_\ell\mathbf{R})$ as the input. The first term in the right hand side of~(\ref{SOFT_METRIC}) specifies the EDF between the candidate codeword and the received vector $\boldsymbol{y}^{(0)}$, while the second term is the EDF between the flipped vector $\boldsymbol{y}^{(1)}$ and its corresponding VA output $\tilde{\boldsymbol{v}}_\ell$. Both of them are likely to be large in the case when the candidate codeword is the transmitted one. Heuristically, we will set a threshold on $M(\hat{\boldsymbol{v}}^{(0)}_\ell)$ to check the correctness of the candidate codeword, as illustrated in \textbf{Example}~\ref{EX_THRESHOLD}.

\begin{figure}
  \centering
  \includegraphics[width=8cm]{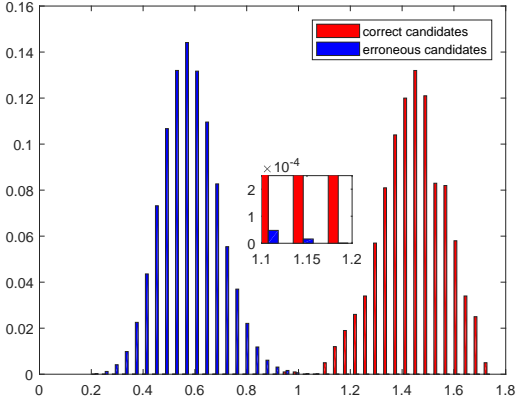}\\
  \caption{Statistical behavior of $M(\hat{\boldsymbol{v}}^{(0)}_\ell)$.}\label{FIG_THRESHOLD}
\end{figure}

\begin{example}\label{EX_THRESHOLD}
The TBCC in \textbf{Example}~\ref{EX_LIST} is taken as the basic code. We set ${\rm SNR} = 3~{\rm dB}$ and $\ell_{\rm max}=64$. The histogram is shown in Fig.~\ref{FIG_THRESHOLD}. We set a threshold $T$ to distinguish the correct decoding candidate from the erroneous one. The decoding candidate $\hat{\boldsymbol{v}}^{(0)}_\ell$ is treated to be correct only if $M(\hat{\boldsymbol{v}}^{(0)}_\ell) \geqslant T$, where $T$ is usually set large~(e.g., $T=1.2$ in this example) to reduce the probability that an erroneous candidate is mistaken as the correct one. The threshold $T$, depending on SNRs and coding parameters, can be learned off-line and stored for use in the decoding algorithm.
\end{example}

The successive cancellation decoding algorithm for BMST-TBCC is outlined as follows. For the first sub-frame, the decoder employs the SLVA to compute the decoding candidates, which will be checked by a statistical threshold, until finding a qualified one. If the list size reaches the maximum $\ell_{\rm max}$ and no decoding candidate is qualified, the decoder delivers $\hat{\boldsymbol{v}}^{(0)}_\ell$ with the maximum $M(\hat{\boldsymbol{v}}^{(0)}_\ell)$ as output. After removing the effect of the first sub-frame, the second sub-frame is then decoded in the same way. This process will be continued until all sub-frames are decoded. The detailed schedule for the decoding algorithm is summarized in Algorithm~\ref{DecodingAlgorithmW2}.

\begin{algorithm}\caption{Successive cancellation decoding for BMST-TBCC}\label{DecodingAlgorithmW2}
\begin{itemize}
  \item{\textbf{Global initialization:}} Set the threshold $T$. Assume that $\boldsymbol{y}^{(0)}$ has been received and set $\boldsymbol{z}^{(0)}=\boldsymbol{y}^{(0)}$.
  \item{\textbf{Sliding-window decoding:}}  For $0\leqslant t \leqslant L-1$, after receiving $\boldsymbol{y}^{(t+1)}$,
  \begin{enumerate}
    \item{\textbf{Local initialization:}} Set $M_{\rm max}=-\infty$ and $\ell = 1$.
    \item{\textbf{List:}}  While $M_{\rm max}\leqslant T$ and $\ell \leqslant \ell_{\rm max}$,
      \begin{enumerate}
        \item Perform SLVA to find $\hat{\boldsymbol{v}}^{(t)}_\ell = {\rm SLVA}(\boldsymbol{z}^{(0)}, \ell)$ and compute $D(\hat{\boldsymbol{v}}^{(t)}_\ell,\boldsymbol{z}^{(0)})$.
        \item Flip the received vector $\boldsymbol{y}^{(t+1)}$, resulting in  $\boldsymbol{z}^{(1)} = \boldsymbol{y}^{(t+1)} \odot \phi(\hat{\boldsymbol{v}}^{(t)}_\ell\mathbf{R})$.
        \item Perform VA to find $\tilde{\boldsymbol{v}}_\ell = {\rm VA}(\boldsymbol{z}^{(1)})$ and compute $D(\tilde{\boldsymbol{v}}_\ell,\boldsymbol{z}^{(1)})$.
        \item If $M(\hat{\boldsymbol{v}}^{(t)}_\ell)=D(\hat{\boldsymbol{v}}^{(t)}_\ell,\boldsymbol{z}^{(0)})+D(\tilde{\boldsymbol{v}}_\ell,\boldsymbol{z}^{(1)}) \geqslant M_{\rm max}$, replace $M_{\rm max}$ by $M(\hat{\boldsymbol{v}}^{(t)}_\ell)$ and $\hat{\boldsymbol{v}}^{(t)}_{\rm max}$ by $\hat{\boldsymbol{v}}^{(t)}_\ell$.
        \item Increment $\ell$ by one.
      \end{enumerate}

    \item{\textbf{Decision:}} Output $\hat{\boldsymbol{u}}^{(t)}$, the corresponding information vector to $\hat{\boldsymbol{v}}^{(t)}_{\rm max}$, as the decoding result of the $t$-th sub-frame.
    \item{\textbf{Cancellation:}} Remove the effect of the $t$-th sub-frame on the $(t+1)$-th sub-frame. That is, update $\boldsymbol{z}^{(0)}$ by computing
    \begin{displaymath}
    \boldsymbol{z}^{(0)} = \boldsymbol{y}^{(t+1)}\odot\phi(\hat{\boldsymbol{v}}^{(t)}_{\rm max}\mathbf{R}).
    \end{displaymath}
  \end{enumerate}
\end{itemize}
\end{algorithm}

\section{Simulation Results}\label{SEC_5}
The $16$-state $(2,1,4)$ TBCC defined by the polynomial generator matrix $G(D) = [10111, 11001]$ is taken as the basic code. The total rate is set to $R = 0.49$ by terminating the codes properly.

\begin{figure}
  \centering
  \includegraphics[width=8cm]{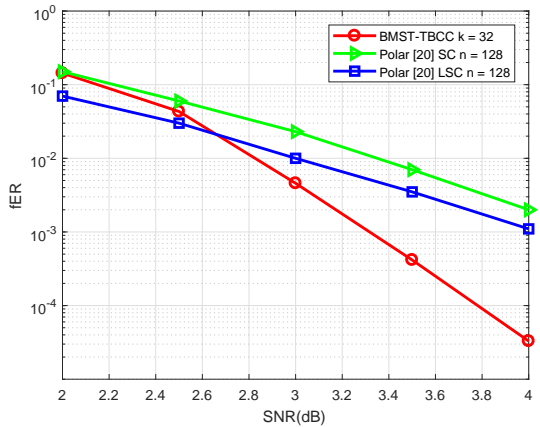}\\
  \caption{Comparison of BMST-TBCC and polar code.}\label{FIG_SIM1}
\end{figure}

\begin{example}\label{EX_1}
We set $k=32$ and $\ell_{\rm max}=64$. A set of thresholds $T_A = 1.3,1.35,1.4,1.45,1.5$ are chosen for ${\rm SNR} = 2.0,2.5,3.0,3.5,4.0$, respectively. The fER is shown in Fig.~\ref{FIG_SIM1}. For comparison, we have also redrawn the performance curves of the polar code~\cite{Wu2016OSDPolar} with length 128~(the same decoding delay as the BMST-TBCC). We observe that the BMST-TBCC with successive cancellation decoding is competitive with the polar code.
\end{example}

\begin{figure}
  \centering
  \includegraphics[width=8cm]{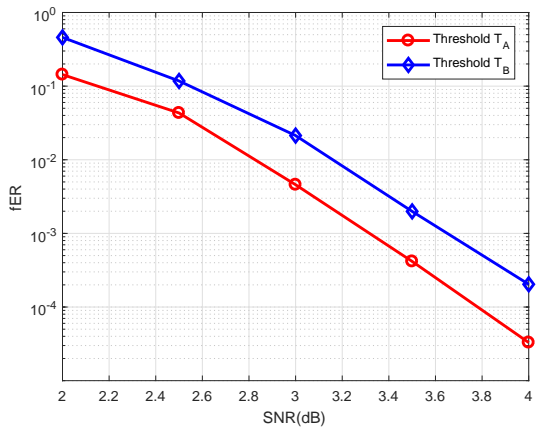}\\
  \caption{Performance of BMST-TBCC with different thresholds.}\label{FIG_SIM2}
\end{figure}

\begin{example}\label{EX_2}
Consider the code in \textbf{Example}~\ref{EX_1} again. Another set of thresholds $T_B = 0.95,1.0,1.05,1.1,1.15$ are chosen for ${\rm SNR} = 2.0,2.5,3.0,3.5,4.0$, respectively. The fER is shown in Fig.~\ref{FIG_SIM2}, while the average list size needed for decoding a sub-frame is shown in Table~\ref{TAB_SIM2}. It can be seen that the complexity~(average list size), at the cost of performance loss, can be reduced by tuning down the threshold. For example, at ${\rm SNR} = 4~{\rm dB}$, the computational complexity~(average list size) can be reduced more than $10$ times if a performance degradation~(fER deterioration) is tolerated from $10^{-5}$ to $10^{-4}$.
\end{example}

\begin{table}[t]
  \centering
  \caption{Average list size needed for $T_A$ and $T_B$}
  \begin{tabular}{|l|l|l|l|l|l|l|}\hline
  ${\rm SNR}$             & 2.0 & 2.5 & 3.0 & 3.5 & 4.0\\\hline
  list size for $T_A$     & 38  & 30  & 23  & 18  & 14 \\\hline
  list size for $T_B$     & 25  & 8.2 & 2.6 & 1.3 & 1.1\\\hline
  \end{tabular}\label{TAB_SIM2}
\end{table}

\section{Conclusion}\label{SEC_6}
In this paper, a new decoding algorithm has been proposed for BMST-TBCC. The decoder outputs serially a list of decoding candidates and identifies the correct one by a statistical threshold, which can be designed by statistical learning and adjusted to make a tradeoff between performance and complexity. Simulation results have been presented to show the performance of the proposed algorithm with different parameters.

\section*{Acknowledgment}
This work was supported by the NSF of China~(No. 61771499) and the Basic Research Project of Guangdong Provincial NSF~(No. 2016A030308008 and No. 2016A030313298).


\bibliographystyle{IEEEtran}
\bibliography{IEEEabrv,bibfile}

\end{document}